\documentclass[aps,preprint,amsmath,amssymb,showpacs,amsfonts,nofootinbib]{revtex4}
\usepackage{epsfig}
\usepackage{graphicx}
\usepackage{dcolumn}
\usepackage{bm}
\usepackage{amsthm}
\usepackage{amsmath}
\usepackage{eufrak}

\newcommand{\beq}{\begin{equation}}
\newcommand{\eeq}{\end{equation}}
\newcommand{\bea}{\begin{eqnarray}}
\newcommand{\eea}{\end{eqnarray}}
\newcommand{\nn}{\nonumber}

\newcommand{\al}{\alpha}
\newcommand{\be}{\beta}
\begin{document}
\begin{flushright}
%AEI-2012-002
\end{flushright}
\bigskip
\bigskip

\title{Constraints on Rindler Hydrodynamics}
\author{Adiel Meyer$^1$}
\email{adielmey@post.tau.ac.il}
\author{Yaron Oz$^1$}
\email{yaronoz@post.tau.ac.il}

\affiliation{School of Physics and Astronomy, Tel Aviv University, Tel Aviv 69978, Israel}

\date{\today}
\begin{abstract}
We study uncharged Rindler hydrodynamics at second order in the derivative expansion. The equation
of state of the theory is given by a vanishing equilibrium energy density.
We derive relations among the transport coefficients by employing two frameworks. First, by the requirement of having an entropy current with a non-negative divergence, second
by studying the thermal partition function on stationary backgrounds. 
The relations derived by these two methods  are equivalent.
We verify the results by studying explicit examples in flat and curved space-time geometries.
 
\end{abstract}

%Physics of black holes, Gauge/string duality,Relativistic fluid dynamics
%\pacs{04.70.-s, 11.25.Tq, 47.10.ad}

\maketitle

\tableofcontents

\newpage

\section{Introduction and Summary}

There has been much interest in recent years in the holographic relation between the hydrodynamics of quantum field
theories defined on $(d+1)$-dimensional space-times and deformations of black hole geometry in one higher space dimension. 
Aspects of this relation have been studied for a large class of field theories in diverse dimensions including conformal and non-confomal theories, relativistic as well as non-relativistic ones, anomalous theories and theories with spontaneously broken symmetries.

One intriguing example, which will be the subject of this paper,  is the Rindler hydrodynamics, that is the hydrodynamics
induced on a codimension one hypersurface in Rindler geometry \cite{Bredberg:2011jq,Compere:2011dx,Compere:2012mt,Eling:2012ni,Eling:2012xa}. The Rindler geometry is a solution to the vacuum Einstein equations, and 
correspondingly the equilibrium energy density in the hypersurface hydrodynamics vanishes.
Despite this unconventional equation of state, the hydrodynamic description is well defined. Viewing hydrodynamics as an effective description of quantum field theory at large length scales and at local thermal equilibrium, one may hope that understanding Rindler hydrodynamics can shed
light on how Holography works in asymptotically flat space-times.

In the hydrodynamic description one upgrades the local charge densities to local currents. In this framework, 
the entropy current in hydrodynamics is an upgrade of the local entropy density to a local current \cite{landau}.
The entropy current has not been given yet a microscopic origin and it is not a unique object. 
It is constructed phenomenologically order by order in the hydrodynamic derivative expansion.
One can generalize the  second law of thermodynamics, by requiring that the entropy current has a non-negative
divergence order by order in this expansion.
The requirement imposes constraints on the transport coefficients of the hydrodynamic description, which in turn impose
constraints on the underlying microscopic quantum field theory. This has been applied, for instance, in the study of anomalous hydrodynamics 
\cite{Son:2009tf,Neiman:2010zi}, uncharged  
hydrodynamics \cite{Bhattacharyya:2012nq}, and  in the analysis of holographic hydrodynamics \cite{Bhattacharyya:2008xc,Chapman:2012my},

There are two types of such constraints. One type gives equality constraints that relate different transport coefficients. The second type
gives inequalities, such as the  requirement
that the shear and bulk viscosities at the first viscous order are non-negative.
In order to fully exploit these constraints one studies them in the presence background fields. In the case of uncharged hydrodynamics the
background field is a curved metric.

A second method to derive the first type of constraints employs the thermal partition function on a curved background with a timelike Killing vector
\cite{Banerjee:2012iz,Jensen:2012jh}.
In this framework the constraints are related to the underlying gauge and diffeomorphism symmetries.

The aim of this paper is to study the first type of  these constraints in the case of Rindler hydrodynamics.
We will employ 
the two methods discussed above and compare the results.
The paper is organized as follows.
In section 2 we will classify the independent fluid data, i.e. modulo the conservation laws of the stress-energy 
tensor in ideal hydrodynamics. We will use as variables the fluid velocity and pressure as well as the curved background data. 
In section 3 we will construct the entropy current up to second order in the derivative expansion and its divergence.
By requiring the latter to be non-negative we will derive constraints on the transport coefficients.
We will check the constraints in two cases: flat hypersurface (with and without a bulk Gauss-Bonnet term) and curved hypersurface in Einstein gravity.
In section 4 we will constrict the thermal partition function and derive the constraints on the transport coefficients.
The constraints obtained by the two methods are equivalent.
Some details of the calculation are outlined in the appendices.

\section{The Fluid Data}

We would like to study the  (d+1)-dimensional Rindler hydrodynamics in a general curved background up to second order in the  derivative expansion.
The fluid variables that we will use are the fluid velocity $u^\mu$ normalized as $u_{\mu}u^{\mu} = -1$, and the fluid pressure $p$. The curvature data will be given by the Riemann tensor $R^\mu_{\nu\rho\sigma}$.
 In Rindler hydrodynamics we have a  vanishing equilibrium 
energy density  $\epsilon_0 = 0$.
We will start by classifying the independent fluid and curvature data, that is, modulo the ideal fluid hydrodynamic equations and the curvature identities.
The classification of fluid and curvature data when  $\epsilon_0 \neq 0$ has been carried out in \cite{Bhattacharyya:2012nq}. One cannot simply use this classification by setting $\epsilon_0 = 0$, since the ideal fluid hydrodynamic equations are different. 
We will use, however, the same classification scheme according to the transformation properties under the local  $SO(d)$ symmetry group that leaves
the (d+1)-velocity $u^{\mu}(x)$ invariant, as employed in  \cite{Bhattacharyya:2012nq}.

The Rindler hydrodynamics stress-energy  tensor at the ideal level is 
\begin{equation}
T_{\mu\nu}^{(0)}=pP_{\mu\nu} \ ,
\end{equation}
 and the conservation equations
 $\nabla_\nu T^{\sigma\nu}=0$ projected on $u_\sigma$ and on $P^\mu_\sigma$ are:
\begin{align}
&\Theta \equiv \nabla_\mu u^\mu = 0 \  ,\\
&u^\sigma \nabla_\sigma u^\mu + P^{\mu\sigma}\nabla_\sigma \text{ln} p = 0  \ .
\end{align}
$P^{\mu\nu} = g^{\mu\nu} + u^\mu u^\nu$ is the  projector on the vector space perpendicular to $u^\mu$.
The fact that the fluid expansion $\Theta$ vanishes will be of importance.

In the following tables we classify all the possible terms after imposing the ideal fluid conservation equations.
The independent fluid data at first order is given in
Table \ref{table:1storder}.
\begin{table}[ht]
\vspace{0.5cm}
\centering % used for centering table
\begin{tabular}{|c| c|} % centered columns (2 columns)
\hline
%\hline %inserts double horizontal lines
 & Independent data \\ [1ex] % inserts table
%heading
%\hline % inserts single horizontal line
\hline
Scalars (1)  & $D \text{ln} p$  \\ [0.5ex]  % inserting body of the table
\hline
Vectors (1) & ${\mathfrak a}^\mu \equiv(u.\nabla)u^\mu$ \\[0.5ex]
\hline
Tensors (2) & $\sigma_{\mu\nu}\equiv  \nabla_{(\mu}u_{\nu)}$ \\
& $\omega_{\mu\nu}\equiv  \nabla_{[\mu}u_{\nu]}$\\[0.5ex]
\hline
%\hline
\end{tabular}
\caption{Fluid data at first order in derivatives} % title of Table
\label{table:1storder} % is used to refer this table in the text
\end{table}

$D = u^{\mu}\nabla_{\mu}$, ${\mathfrak a}_\mu$ is the fluid acceleration, $\sigma^{\mu\nu}$ and $\omega^{\mu\nu}$ are the fluid shear and rotation tensors, respectively.
We define $A_{(\mu}B_{\nu)} = \frac{1}{2} (A_\mu B_\nu+A_\nu B_\mu )$ and  $A_{[\mu}B_{\nu]} = \frac{1}{2} [A_\mu B_\nu-A_\nu B_\mu ]$ .\\
At this order there is no curvature data.
The independent fluid data at second order is derivatives in given  in Table \ref{table:2ndorder}.
\begin{table}[ht]
\vspace{0.5cm}
\centering % used for centering table
\begin{tabular}{|c| c|} % centered columns (2 columns)
\hline %inserts double horizontal lines
 &Independent data \\ [1ex] % inserts table
%heading
\hline % inserts single horizontal line
Scalars (1) &  $ D(D\text{ln} p)$ \\ [0.5ex]  % inserting body of the table
\hline
Vectors (2)&  $P^{\mu \al}\nabla_{\al} (D \text{ln} p)$ \\
&$P^\mu_{\al}\nabla_{\be}\sigma^{\al\be}$\\[0.5ex]
\hline
Tensors (1)& $P^{\mu \al}P^{\nu \be} \nabla_{\al} \nabla_{\be}\text{ln} p$ \\[0.5ex]
\hline
\end{tabular}
\caption{Fluid data at second order in derivatives} % title of Table
\label{table:2ndorder} % is used to refer this table in the text
\end{table}
The composite expressions of fluid data that we will need for the analysis are given in Table \ref{table:c11}.
\begin{table}[ht]
\vspace{0.5cm}
\centering % used for centering table
\begin{tabular}{|c| c|} % centered columns (2 columns)
\hline%inserts double horizontal lines
Scalars (4) &$(D\text{ln} p)^2,~~{\mathfrak a}^2,~~\omega^2 ,~~\sigma^2$
\\[0.5ex]
\hline
Vectors (3) &${\mathfrak a}^\mu D\text{ln} p,~~{\mathfrak a}_\nu \omega^{\mu\nu},
~~{\mathfrak a}_\nu \sigma^{\mu\nu}$\\[0.5ex]
\hline
Tensors (5) &$D\text{ln} p \sigma_{\mu\nu}$,
~ $\sigma_{\langle\mu}^{\al}\sigma_{\al\nu\rangle}$,~$\omega_{\langle\mu}^{\al}\sigma_{\al\nu\rangle}$,~
$\omega_{\langle\mu}^{\al}\omega_{\al\nu\rangle}$,
~${\mathfrak a}_{\langle \mu}{\mathfrak a}_{\nu\rangle}$\\[0.5ex]
\hline
\end{tabular}
\caption{Composite expressions in fluid data at second order} % title of Table
\label{table:c11} % is used to refer this table in the text
\end{table}
We define 
\begin{equation}
A_{\langle\mu\nu\rangle} \equiv P_\mu^\alpha P_\nu^\beta\left(\frac{A_{\alpha\beta} + A_{\beta\alpha}}{2} - g_{\alpha\beta}
\left(\frac{P^{\rho\sigma}A_{\rho\sigma}}{d}\right) \right) \ .
\end{equation}
where $\omega^2 =\omega_{\mu\nu}\omega^{\mu\nu} $
The classification of the independent curvature data is the same as in  \cite{Bhattacharyya:2012nq} and for completeness we give it in
table \ref{table:2ndcurv}.

\begin{table}[ht] 
\vspace{0.5cm}
\centering % used for centering table
\begin{tabular}{|c| c| } % centered columns (4 columns)
\hline
 & Independent data\\[0.5ex]
\hline
 &  $R\equiv {R^{\mu\nu}}_{\mu\nu}$\\
Scalars (2)&$R_{00}\equiv u^\mu u^\nu R_{\mu\nu}$\\
&$~~~~~~~\equiv u^\mu u^\nu {R^\alpha}_{\mu\alpha\nu}$\\ [0.5ex]  % inserting body of the table
\hline
Vectors(1) &$ P^{\mu \al}R_{\al\be} u^\be$\\[0.5ex]
\hline
& $R_{\langle \mu\nu\rangle}$\\
Tensors(2) &$K_{\langle\mu\nu\rangle}$\\
& where
 $K_{\mu\nu} \equiv u^\alpha u^\beta R_{\mu\alpha\nu\beta}$\\[0.5ex]
\hline
\end{tabular}
\caption{Independent type curvature data at second order} % title of Table
\label{table:2ndcurv} % is used to refer this table in the text
\end{table}

Consider next the the most general form of the stress-energy tensor of Rindler hydrodynamics in curved space. 
We fix the ambiguities at the derivative order by requiring that \cite{Eling:2012ni,Compere:2012mt,Eling:2012xa}
\beq
P^{\mu}_{\sigma}T^{(n)}_{\mu\nu}u^{\nu} = 0 \ ,
\eeq
and the pressure receives no derivative corrections. With this choice,
the stress-energy tensor up to second order in derivatives takes the form

\begin{equation} \label{cru}\begin{split}
T^{(0)}_{\mu\nu}+\Pi_{\mu\nu} =~& pP_{\mu\nu}-2\eta\sigma_{\mu\nu} - \zeta u_\mu u_\nu D \text{ln} p\\
~&+p^{-1}\bigg[ \tau ~\nabla_{<\mu}\nabla_{\nu>}\text{ln} p + \kappa_1 R_{<\mu\nu>} + \kappa_2 K_{ \mu\nu} +\lambda_0~D \text{ln} p \sigma_{\mu\nu}\\
&+ \lambda_1~ {\sigma_{\mu}}^{\al}\sigma_{\al\nu}+ \lambda_2~ {\sigma_{( \mu}}^{\al}\omega_{\al\nu )}+ \lambda_3~ {\omega_{ \mu}}^{\al}\omega_{\al\nu} + \lambda_4~{\mathfrak a}_{\mu}{\mathfrak a}_{\nu}\bigg]\\
&+u_\mu u_\nu\bigg[ d_1 \sigma^2 - d_2 \omega^2
 + d_3 (D \text{ln} p)^2 +d_4 D(D\text{ln} p) +d_5{\mathfrak a}^2+ e_1 R + e_2 R_{00} 
 \bigg] 
\end{split} 
\end{equation}
where
\begin{equation}
K^{\mu\nu} =  R^{\mu \alpha \nu \beta}u_{\alpha} u_{\beta},~~ R^{\mu\nu}
= R^{\alpha\nu \beta\nu}g_{\alpha\beta} \ . 
\end{equation}
There are two transport coefficients at first order: $\eta$ is the shear viscosity and $\zeta$ is a contribution to the energy density. There are fifteen transport coefficients at second order. This is the same number as
for a generic non-conformal hydrodynamics with non-vanishing equilibrium energy density. However, some of the transport coefficients
multiply different expressions in the fluid variables. In particular, seven of these terms can be viewed as corrections to the energy density.
As we saw in the previous section, the fluid expansion $\Theta = \nabla_\mu u^\mu$ vanishes and the scalar $Dlnp$ appears instead in the stress-energy 
tensor.
Note, however, that unlike the case where the equilibrium energy density is nonzero, here  this scalar is not
 related to $\Theta$.

\section{The Entropy current}

In this section we will construct the most general entropy current that exhibits vanishing equilibrium energy density. We will construct it up to second order in derivatives and derive the constraints on its form and on the transport coefficients in the stress-energy tensor. These follow from the requirement that its divergence is non-negative.
 
\subsection{The entropy current at second order}

In \cite{Bhattacharyya:2012nq}, the most general entropy current when the equilibrium energy density is nonzero has been constructed. 
In the following we will use a similar procedure and outline the differences.
As above, we will use as fluid variables the normalized fluid velocity $u^\mu$ and the pressure $p$, and
replace the fluid expansion $\Theta$ by the scalar $Dlnp$.

The entropy current $J^{\mu}$ has a derivative expansion 
\begin{equation}
J^{\mu} = \sum_{l\geq 0} J^{\mu}_{(l)} \ ,
\end{equation}
where $J^{\mu}_{(0)} = s u^{\mu}$ and $J^{\mu}_{(1)} = a_1(p)u^\mu D \text{ln} p + a_2(p)\mathfrak{a}^\mu$.
At second order in derivatives we get that the most general entropy current depends on thirteen parameters, which correspond to the six independent
vectors (three of them are composite), and
seven independent scalars (multiplied by $u^{\mu}$) at second order, as outlined in the tables. 
It is convenient to parametrize the entropy current at the second order as follows:
\begin{equation}\label{duient1}
 \begin{split}
 J^\mu_{(2)} =& \nabla_\nu\left[A_1(u^\mu\nabla^\nu \text{ln} p  - u^\nu \nabla^\mu \text{ln} p)\right] + \nabla_\nu \left( A_2 \text{ln} p \  \omega^{\mu\nu}\right)\\
 & + A_3 \left(R^{\mu\nu} - \frac{1}{2}g^{\mu\nu} R\right) u_\nu
+\left[ A_4 D (D\text{ln} p) + A_5 R + A_6 R_{00}\right] u^\mu\\ &+(B_1\omega^2 + B_2 (D \text{ln} p)^2 + B_3 \sigma^2)u^\mu + B_4\left[\mathfrak{a}^2
u^\mu - 2 (D \text{ln} p) \nabla^\mu_\perp \text{ln} p\right]\\
&+\left[D \text{ln} p  \nabla^\mu B_5 - P^{\al\be}(\nabla_\be u^\mu)( \nabla_\al B_5 )\right]+ B_6 D\text{ln} p  {\mathfrak a^\mu} + B_7 {\mathfrak a}_\nu \sigma^{\mu\nu} \ .
 \end{split}
 \end{equation}

The divergence of the entropy current up to second order reads \eqref{duient1} :
\begin{align}\label{divergence}
\nabla_\mu J^\mu &=\nabla_\mu (s u^\mu)+ \nabla_\mu J^\mu_{(1)}+ \nabla_\mu J^\mu_{(2)} \nonumber\\
&=-\frac{s}{p}\sigma_{\mu\nu}\Pi^{\mu\nu}-\frac{s}{p}D\epsilon + \nabla_\mu J^\mu_{(1)} +\nabla_\mu J^\mu_{(2)} \geq 0 \ , 
\end{align}
where the energy density upto second order $\epsilon = T_{\mu\nu}^{(0+1+2)}u^{\mu}u^{\nu}$ reads
\begin{align}
\epsilon &= -\zeta D \text{ln} p + d_1\sigma^2-d_2\omega^2+d_3(D\text{ln} p)^2+d_4D^2\text{ln} p+d_5 \mathfrak{a}^2+e_1R+e_2R_{00} \ .
\end{align}
Note, that the entropy density $s$ is constant at zeroth order since it is related to the energy density by the thermodynamic relation $d\epsilon = T ds$. 
\subsection{Constraints on the entropy current and the transport coefficients}
In order to constrain the entropy current and to find relations among the transport coefficients in the stress-energy tensor, we have to analyze the terms that can appear in the divergence of the entropy current. We omit in the entropy divergence terms that cannot be completed to a positive expression, that is, to a complete square.
We will encounter two cases.
In the first case, we find terms that cannot appear in the second order divergence of the entropy current, i.e.  $\mathfrak{a}^2$.
We will therefore set to zero the coefficients of all the third order terms in the entropy current divergence that take the form of second order vectors times $\mathfrak{a}^\mu$. 
In the second case, we identify terms that cannot appear at the fourth order divergence. This analysis was carried out in \cite{Bhattacharyya:2012nq} for non-vanishing energy density. We can use this analysis with the following modifications: 
\begin{itemize}
\item 
In \cite{Bhattacharyya:2012nq} change $\Theta$ to $D \text{ln} p$.
\item 
In order to use the basis used in \cite{Bhattacharyya:2012nq}, we compute the entropy divergence (\ref{divergence2}) and change the basis from $P^{\mu\alpha}P^{\nu\beta}\nabla_\alpha\nabla_\beta \text{ln} p$ to $ P^{\mu\alpha}P^{\nu\beta}D\sigma_{\alpha\beta} $ by,
\begin{align}
P^{\mu\alpha}P^{\nu\beta}\nabla_\alpha\nabla_\beta \text{ln} p =\mathfrak{ a^\mu a^\nu} + \sigma^{\mu\nu} D \text{ln} p  - \sigma^{\mu \rho} \sigma_\rho^{~\nu} -  \omega^{\mu \rho} \omega_\rho^{~\nu} - P^{\mu\alpha}P^{\nu\beta}D\sigma_{\alpha\beta} - K^{\mu\nu} .
\end{align}
\end{itemize}
Therefore, the following terms cannot appear  at the fourth order divergence: 
$(\omega)^4,~ (\omega.a)^2,~ (a)^4,~ R^2,~ R_{00}^2,~ R_{\mu\nu}R^{\mu\nu},~  K_{\mu\nu}K^{\mu\nu} $ .
We will, thus, set to zero the coefficients of all the third order terms in the entropy current divergence that cannot be organized in a complete square, i.e. 
\begin{align}
&D\text{ln} p R_{00}, ~ RD\text{ln} p,~ R^{\mu\nu}\sigma_{\mu\nu},~R^{\mu\nu} \mathfrak{a}_\nu u_\mu,~ \mathfrak{a}^2D\text{ln} p ,~D^3\text{ln} p, ~ \mathfrak{a}^\beta \nabla_\beta(D\text{ln} p),\nonumber\\
&\sigma^{\beta \mu} \mathfrak{a}_\beta \mathfrak{a}_\mu,~DR,~ DR_{00},~\omega^2D \text{ln} p ,~\sigma^{\nu \lambda} K_{\nu \lambda},~\sigma^{\nu \lambda}\omega_\lambda^{~ \sigma}\omega_{\sigma \nu},~ \mathfrak{a}_\nu \nabla_\mu \sigma^{\mu\nu}. \nonumber
\end{align}

For this we will have to construct the most general entropy current to third order in derivatives and take its divergence.

Carrying out this analysis, we find the most general entropy current with positive divergence:
\begin{equation}\label{duient2}
 \begin{split}
J^\mu_{(1)} =& -\frac{s}{p} \zeta D \text{ln} p \\
 J^\mu_{(2)} =& \nabla_\nu\left[A_1(u^\mu\nabla^\nu \text{ln} p  - u^\nu \nabla^\mu \text{ln} p)\right] + \nabla_\nu \left( A_2 \text{ln} p \  \omega^{\mu\nu}\right)\\
 & + A_3 \left(R^{\mu\nu} - \frac{1}{2}g^{\mu\nu} R\right) u_\nu
+\left[ A_4 D (D\text{ln} p) -\frac{p}{2}\frac{dA_3}{dp} R + A_6 R_{00}\right] u^\mu\\ &+\left(B_1\omega^2 + B_2 (D  \text{ln} p)^2 + B_3 \sigma^2\right)u^\mu \\
&- \left(p\frac{dA_3}{dp} + \frac{p^2}{2}\frac{d^2A_3}{dp^2} + \frac{1}{2}A_6 + \frac{p}{2}\frac{dA_6}{dp} \right)\left[\mathfrak{a}^2
u^\mu - 2 (D\text{ln} p) \nabla^\mu_\perp \text{ln} p\right]\\
&+\left(A_3 + p\frac{dA_3}{dp}\right)\left[D\text{ln} p  \nabla^\mu \text{ln} p - P^{\al\be}(\nabla_\be u^\mu)( \nabla_\al \text{ln} p )\right]\\
&+\left( A_3+4p\frac{dA_3}{dp} + p^2\frac{d^2A_3}{dp^2} + 2A_6 + p\frac{dA_6}{dp} \right)D \text{ln} p   {\mathfrak a^\mu} \end{split}
 \end{equation}
with,
\begin{align}
&A_3 = p^{-2}\kappa_1,~~ A_4 =p^{-1} d_4,~~  A_6 =p^{-1} e_2,~~ B_1 = \frac{1}{4}\left(A_3+p\frac{dA_3}{dp} + p^{-2}(\tau-\lambda_3-4d_2)\right)
\end{align}
which leave four free parameters, $A_1,~A_2,~B_2,~ B_3$. For a detailed computation of its divergence see Appendix \ref{entropy div}.
 
 The relations among the stress-energy tensor transport coefficients that we get are:
\begin{align}
\kappa_2 &= \tau -\kappa_1+p\frac{d\kappa_1}{dp} \label{relation1}\\
pe_1  &= \kappa_1 - \frac{p}{2}\frac{d\kappa_1}{dp} \label{relation2} \\
\lambda_4 &= p(2e_1-e_2) + p\frac{d}{dp}(\tau + \kappa_1 - \kappa_2) - 2 \tau - \kappa_1 + \kappa_2  \label{relation3} \\
0& = pd_2 +p e_1 +p e_2 -2\kappa_1 - \kappa_2 + \lambda_3 + \frac{p}{4} \frac{d}{dp}( 2\kappa_1 + \kappa_2 - \lambda_3)  \label{relation4} \\
pd_5 &= pe_1 - pe_2 + p^2\frac{d}{dp}(2e_1 - e_2)  \label{relation5} \ .
\end{align}

\subsection{Flat space-time background}
In the particular case of hydrodynamics in flat space-time, we set to zero the curvature terms in the entropy current and in the stress-energy tensor. This restrict the coefficients of the entropy current and we get the relations:
\begin{align}\label{flatD1}
A_1+A_3=A_6,~~A_2+A_3=0,~~A_3=2A_5 .
\end{align}
The relations among the coefficients of the entropy current that follow from the requirement of non-negative divergence of the entropy current are,

\begin{align}
&4B_1+\frac{4d_2}{p}+\frac{\lambda_3}{p^2} - p\frac{dB_5}{dp} - \frac{\tau}{p^2} =0 \label{FlatC1} \\
&-2B_4 - p^2 \frac{d^2B_5}{dp^2}+\frac{2d_5}{p} - \frac{\lambda_4}{p^2} - p\frac{dB_5}{dp} -\frac{\tau}{p^2} = 0 \\
& - p\frac{dB_4}{dp} - 2B_4 + p^2 \frac{d^2B_5}{dp^2} + p\frac{dB_5}{dp} -  p\frac{dB_6}{dp} + \frac{2d_5}{p} - \frac{dd_5}{dp} = 0 
\end{align}
\begin{align}
& A_4 = \frac{d_4}{p} \\
&  p\frac{dB_5}{dp} = B_6 + \frac{2d_5}{p} \\
& p\frac{dB_1}{dp} - 2B_1 + 2B_4 - 2 p\frac{dB_5}{dp} + B_6 + \frac{dd_2}{dp} - \frac{2d_2}{p} = 0 \\
&B_7 = 0 \label{FlatC8} \ ,
\end{align}
We can eliminate the coefficients of the entropy current from (\ref{FlatC1}-\ref{FlatC8}) to get a relation between the transport coefficients of the stress-energy tensor,
\begin{align}\label{FlatT}
&\frac{4d_2}{p}-4\frac{dd_2}{dp}+\frac{8\lambda_3}{p^2} - \frac{5}{p}\frac{d\lambda_3}{dp}+\frac{d^2\lambda_3}{dp^2} - \frac{8\lambda_4}{p^2} +\frac{1}{p}\frac{d\lambda_4}{dp}-\frac{16\tau}{p^2}
+ \frac{6}{p}\frac{d\tau}{dp} - \frac{d^2 \tau}{dp^2} + \frac{6d_5}{p}=0 \ .
\end{align} 
The relations (\ref{FlatC1}-\ref{FlatC8}) can be checked against two examples of Rindler hydrodynamics derived
via the gravitational solutions in one higher space dimension.
The first example is the fluid hydrodynamics obtained in the framework of Einstein's gravity, i.e. from a gravitational background
obeying the Einstein equations for an empty universe $\bold{R}_{AB} = 0$. This example was studied in \cite{Eling:2012ni,Compere:2012mt}, and its results were:
\begin{align}\label{example1}
&A_1=2sp^{-2},~~A_2=-A_3=sp^{-2},~~A_4=0,~~A_5=-\frac{1}{2}sp^{-2},~~A_6=sp^{-2},\nonumber\\
&B_1=\frac{1}{2}sp^{-2},~~B_2=B_4=B_6=2s p^{-2},~~B_3=3sp^{-2},~~p\frac{dB_5}{dp}=2sp^{-2},~~B_7=0 \nonumber\\
&d_1=-2p^{-1},~~d_2=d_3=d_4=d_5=e_1=e_2=0, \nonumber\\
&\tau = \lambda_0 = \lambda_2 = \lambda_3 = -\lambda_4 = -4,~~ \lambda_1 = -2 \nonumber\\
&\eta = 1,~~ \zeta = 0 \ .
\end{align}
$s=4\pi$ is the entropy density in equilibrium in this case.\\
The second example is an empty universe in Gauss-Bonnet gravity. This case has been worked out in \cite{Eling:2012xa}, its results are the same as in the previous example with the following modifications:
\begin{align}\label{example2}
&A_2=-A_3=sp^{-2}(1+4\alpha p^2),~~A_5=-\frac{1}{2}sp^{-2}(1+4\alpha p^2),~~A_6=sp^{-2}(1-4\alpha p^2), \nonumber\\
&B_1=\frac{1}{2}sp^{-2}(1+4\alpha p^2),~~B_2=B_6=2s p^{-2}(1-2\alpha p^2),~~B_3=sp^{-2}(3+2\alpha p^2), \nonumber \\
&p\frac{dB_5}{dp}=2sp^{-2}(1-2\alpha p^2), \nonumber\\
&\lambda_3 = -4(1+3\alpha p^2),~~ \lambda_1 = -2(1+2\alpha p^2) \ ,
\end{align}
where $\alpha$ is the coefficient of the Gauss-Bonnet term. In both cases the relations (\ref{flatD1}-\ref{FlatC8}) are satisfied. Therefore, the relation (\ref{FlatT}) is satisfied as well.\\
Consider  now the special case $\zeta = 0$. We get  two additional constraints:
\begin{align}
& p\frac{A_4}{dp} + 2B_2 - 2p\frac{dB_5}{dp} - \frac{2d_3}{p} - \frac{dd_4}{dp} =0 \\
&  p^2 \frac{d^2B_5}{dp^2} + p\frac{dB_5}{dp} - p\frac{dB_2}{dp} + \frac{dd_3}{dp} = 0 \ .
\end{align}
The case $\zeta = 0$ in flat space has been studied in the holographic framework in \cite{Compere:2012mt}, where it was found that
\begin{align} \label{FlatRestrictions}
&d_2=d_3=d_4=d_5=0,~~ \tau = \lambda_3 = -\lambda_4,~~ B_i=\tilde{B}_i p^{-2} \ ,
\end{align}
with $i=1...7$ and $\tilde{B}_i$ constants. Repeating the analysis for this case, and 
evaluating the constraints for (\ref{FlatRestrictions}) we get the linearly independent equations 
\begin{align}
& \tilde{B}_2 + 2 \tilde{B}_5 =0 \\
& A_4 = 0 \\
& 2\tilde{B}_5 + \tilde{B}_6 =0 \\
&2\tilde{B}_1 - \tilde{B}_4 -\tilde{B}_5  = 0 \\
&\tilde{B}_7 = 0  \\
& \tilde{B}_2-\tilde{B}_4 = 0 \ . \label{additionalcon}
\end{align}
We find six constraints instead of the five that were found in \cite{Compere:2012mt}. Indeed, the higher order entropy current yields one more constraint (\ref{additionalcon}).

\subsection{Curved space-time background}
In this subsection we will construct a solution in Einstein gravity  that describes holographically
Rindler hydrodynamics in a curved background. The additional space coordinates will be denoted by $r$ and the hydrodynamics
is defined on the co-dimension one hypersurface $r=r_c$,
We will follow the procedure in  \cite{Compere:2011dx,Compere:2012mt,Eling:2012ni}, where a flat metric $\eta_{\mu\nu}$ was chosen on $r=r_c$, and we will replace
it by a curved metric $g_{\mu\nu}$.
The details of the construction are presented in appendix \ref{Curved background solution}.
We outline them in the following.
We need to construct the bulk metric up to second order in derivative expansion. We do that by solving the bulk equations $\bf{R}_{AB}=0$, requiring regularity of the solution at $r=0$ and fixing the metric on the hypersurface  $r=r_c$. The difference compared to \cite{Compere:2012mt,Eling:2012ni} appears at second order in the derivative expansion, because curvature effects are of second order or higher. 
The second order correction to the bulk metric reads
\begin{align} \label{curvedmetricC}
\delta_{curv} G_{\mu\nu}^{(2)}&= 2u_\mu u_\nu (r-r_c)R - 2 p^2(r-r_c)^2 u_{(\mu} P^\rho _{\nu)}R^\lambda_\rho u_\lambda + 2(r-r_c)(R_{\mu\nu}+K_{\mu\nu})  \ ,
\end{align}
where $R$, $R_{\mu\nu}$ and $K_{\mu\nu}$ refer to the curvature of the hyper surface metric.
The entropy (area) current  $J^\mu = \frac{1}{4G_{N}}\sqrt{G}\ell^\mu$, where $G_N$ is the Newton constant, is affected by 
the corrections of the metric determinant and the normal to the event horizon:
\begin{align}
\delta_{curv} G^{(2)}&= -\frac{2}{p}(R + R_{00}) \ , \\
\delta_{curv} \ell^\mu_{(2)}&= -\frac{1}{p^3}P^{\mu \rho} R^\lambda_\rho u_\lambda  \ .
\end{align}

The final result is the same as (\ref{example1}) with the following modifications:
\begin{align}\label{example3}
&A_3=A_5=-2sp^{-2},~~e_1=-2p^{-1},~~\kappa_1=\kappa_2 = -2  \ .
\end{align}
It is easy to see that this result satisfies the constraint equations: (\ref{relation1}-\ref{relation5}).\\

 \section{Thermal Partition Function}\label{sec:finalconstraint}

In this section we will consider the Rindler thermal partition function on curved backgrounds with a time-like killing vector  
\cite{Banerjee:2012iz,Jensen:2012jh}.
We will use the framework employed in \cite{Banerjee:2012iz} to find relations among the transport coefficients 
for uncharged non-conformal hydrodynamics. As above, since the equilibrium energy density vanishes in our case, we cannot
simply borrow the results of  \cite{Banerjee:2012iz}  that assume its non-vanishing.
In the case of non-vanishing energy density, the relations among the transport coefficients derived in \cite{Banerjee:2012iz} are the same as those obtained from the non-negativity of the entropy current divergence \cite{Bhattacharyya:2012nq}. We will see this in here as well.

The equilibrium backgrounds posses a time-like killing vector and can be written in the 
Kaluze-Klein form
\begin{align}
ds_{KK}^2=G_{\mu \nu} dx^\mu dx^\nu 
=-e^{2 \sigma(\vec x)} \left( dt+ a_{i}(\vec x) dx^i \right)^2
+g_{ij}({\vec x}) dx^i dx^j \label{ansatz} \ ,
\end{align}
where $i = 1 . . . d$.\\
The fluid variables take the form
\begin{eqnarray}
u^\mu_0 &=&  e^{-\sigma(\vec{x})}(1,0,...,0)  \ , \nonumber\\
p &=& p_0 e^{-\sigma(\vec{x})} \nonumber\\
\epsilon_0 &=& 0 \  .
\end{eqnarray}
\subsection{The stress-energy tensor}
Evaluating the derivative part of the stress-energy tensor \eqref{cru} on the curved equilibrium background we get
\begin{eqnarray}\label{gipi}
\Pi_{eq}^{ij}&=&a_1\bigg(R_g^{ij}-\frac{R_g}{2}g^{ij}\bigg)+a_2\bigg(\nabla^i\nabla^j \sigma-\nabla^2 \sigma g^{ij}\bigg)
+a_3\bigg(\nabla^i \sigma \nabla^j \sigma-\frac{(\nabla \sigma)^2}{2}g^{ij}\bigg)\nn\\
&+&a_4 \bigg(f^{i k}f_{k}^{~j}+\frac{f^2}{4}g^{ij}\bigg)e^{2\sigma}
+g^{ij}\bigg(b_1 R_g + b_2 \nabla^2 \sigma + b_3 (\nabla \sigma)^2+b_4 \frac{1}{4} f^2e^{2\sigma}\bigg) \\
~~\mbox{where}, \nn \\
f_{ij} &=& \partial_i a_j - \partial_j a_i, \quad f^2 = f_{ij}f^{ji} \nonumber\\
{b_1}{p}&=& \frac{1}{2}\kappa_1, \quad {b_2}{p}=\kappa_2- \kappa_1 - \tau \nn \\
{b_3}{p}&=&\frac{1}{2}( \kappa_2 - \kappa_1+ \lambda_4) \nn \\
{b_4}{p}&=&  \frac{1}{4}(2 \kappa_1 + \kappa_2 - \lambda_3) \ ,
\quad {a_1}{p}=\kappa_1 \nn \\
{a_2}{p}&=&\kappa_2-\kappa_1 - \tau, \quad {a_3}{p}=\kappa_2-\kappa_1+\lambda_4, \quad {a_4}{p}=-\frac{1}{4}(2\kappa_1+ \kappa_2) + \frac{1}{4} \lambda_3  \ .
\end{eqnarray}
$R_g^{ij}, R_g$ are Ricci tensor and Ricci scalar respectively, calculated from the spatial metric $g_{ij}(\vec{x})$ in (\ref{ansatz}). The corrections of the fluid variables to second order are:
\begin{equation}\begin{split} \label{vcuf}
u^{\mu}&= b_0u_0^{\mu} +  v_1 e^\sigma \nabla_j \sigma f^{ji} + v_2 e^\sigma \nabla_jf^{ji}   ,   \nn \\
b_0 &=1 - (v_1  e^{2\sigma} a_i\nabla_j \sigma f^{ji} + v_2 e^{2\sigma} a_i\nabla_jf^{ji})\nn \\
p&= p_0 e^{-\sigma} +  t_1 R_g + t_2 \nabla^2 \sigma + t_3 (\nabla \sigma)^2 + t_4\frac{1}{4} f^2 e^{2\sigma}  \\
\epsilon&=   r_1 R_g + r_2 \nabla^2 \sigma + r_3 (\nabla \sigma)^2 +  r_4\frac{1}{4} f^2 e^{2\sigma} \ ,
\end{split}
\end{equation}
where
\begin{equation}
r_1 = e_1,~~r_2 =e_2-e_1 ,~~r_3 =e_2-e_1 +d_5 ,~~r_4 =-(2e_1 + e_2 +d_2) \  .
\end{equation}
The second source of corrections arises from inserting the velocity and pressure corrections into the 
zeroth order stress-energy tensor. We find that the
modification of the stress-energy tensor due to these corrections
is given by
\begin{equation}\label{ccvmu} \begin{split}
T^{ij}&=  g^{ij} \left( t_1 R_g + t_2 \nabla^2 \sigma + t_3 (\nabla \sigma)^2 + t_4 \frac{1}{4} f^2 e^{2\sigma} \right) \\
T_{00}&=  e^{2\sigma} \left( r_1 R_g + r_2 \nabla^2 \sigma + r_3 (\nabla \sigma)^2 +  r_4\frac{1}{4} f^2 e^{2\sigma} \right) \\
T_0^i&=-p_0  \left( v_1 e^\sigma \nabla_j \sigma f^{ji} + v_2 e^\sigma \nabla_jf^{ji}  \right)
\end{split}
\end{equation}
Adding  \eqref{gipi} to  \eqref{ccvmu} we get
\begin{equation}\label{ccvmfu} \begin{split}
T^{ij}&=   g^{ij} \left( t_1 R_g + t_2 \nabla^2 \sigma + t_3 (\nabla \sigma)^2 + t_4 \frac{1}{4} f^2 e^{2\sigma}\right) + \Pi_{eq}^{ij}  \\
T_{00}&=  e^{2\sigma} \left( r_1 R_g + r_2 \nabla^2 \sigma + r_3 (\nabla \sigma)^2 +  r_4\frac{1}{4} f^2 e^{2\sigma} \right)   \\
T_0^i&=-p_0  \left( v_1 e^\sigma \nabla_j \sigma f^{ji} + v_2 e^\sigma \nabla_jf^{ji}  \right)
\end{split}
\end{equation}
where $\Pi^{ij}_{eq}$ was listed in \eqref{gipi}.

\subsection{The partition function}

The thermal partition function takes the form
\begin{equation}\label{cfacnu}
\begin{split}
W = \log Z &= -\frac{1}{2}\int d^dx~\sqrt{g_d} \left[ S_1 R_g + p_0^2 S_2  f_{ij}f^{ij} 
+ S_3 (\nabla\sigma)^2\right]  \ , \\
\end{split}
\end{equation}
where $S_1,S_2,S_3$ are three arbitrary function of $\sigma$.

In order to evaluate the stress-energy tensor from the partition function, we vary the partition function
\begin{eqnarray}\label{variation}
\delta W &=& \int d^{d+1}x \sqrt{-G_{d+1}}\left(-\frac{1}{2}T_{\mu\nu} \delta g^{\mu\nu}\right)= \frac{s}{p_0} \int d^dx \sqrt{-G_{d+1}} \left(-\frac{1}{2}T_{\mu\nu} \delta g^{\mu\nu}\right) \ , 
\end{eqnarray}
where we integrated over the time coordinate  $t$ which is compactified on a circle with radius $\frac{p_0}{s}$. One obtains the stress-energy tensor by varying with respect to the metric,
\begin{eqnarray}\label{variation}
T_{00}&=& -\frac{p_0 e^{2 \sigma}}{s\sqrt{-G_{(d+1)}} }\frac{\delta W}{\delta \sigma},
\quad T_0^i= \frac{p_0}{s\sqrt{-G_{(d+1)}} }\frac{\delta W}{\delta a_i} , \nonumber \\
T^{ij}&=& -\frac{2p_0}{s\sqrt {-G_{(d+1)}}} g^{il}g^{jm}\frac{\delta W}
{\delta g^{lm}}.
\end{eqnarray}

The stress-energy tensor derived from the thermal partition functions (\ref{cfacnu}) reads
\begin{eqnarray}\label{stcuchactionu}
sT^{ij}&=& p S_1 \big(R_g^{ij}-\frac{1}{2}R_g g^{ij}\big)+ 2 p_0^2 p S_2 \big(f^{ik}f_{jk}-\frac{1}{4}f^2 g^{ij}\big) +p
(S_3-S_1'')\big(\nabla^i \sigma\nabla^j\sigma \nn \\
&-&\frac{1}{2}(\nabla \sigma)^2g^{ij}\big)-p S_1' \big(\nabla^i\nabla^j \sigma-g^{ij}\nabla^2\sigma \big) + \frac{1}{2}p S_1''(\nabla \sigma)^2
g^{ij} \nn \\
sT_{00}&=& \frac{p_0^2} {2p}\big(S_1' R_g + p_0^2 S_2' f^2 - S_3' (\nabla \sigma)^2- 2 S_3\nabla^2 \sigma)\big) \nn \\
sT_0^i&=&2 p_{0}^2 p\big(S_2'\nabla_j\sigma f^{ji}+ S_2 \nabla_j f^{ji}\big),
\end{eqnarray}where $'$ denotes derivative with respect to $\sigma$.

Comparing the equations (\ref{ccvmfu})  and  (\ref{stcuchactionu}), one can express the
transport coefficients in terms of the  three coefficients $S$ appearing in (\ref{cfacnu}).
We find
\begin{equation}\begin{split}
  sa_1&= p S_1,~~ sa_{2} = -p S'_1 \quad sa_4= 2 p_0^2p S_2, \quad sa_3= p (S_3- S_1'') \\
  sb_1&=-p t_1,~~ sb_2 = -p t_2,~~ sb_3 = -p t_3 + \frac{1}{2}pS_1'',~~ sb_4 = -p t_4 .
\end{split}
\label{Intrel1}
 \end{equation}
One can eliminate the coefficients $S's$ from above set of relations which gives a relation among transport coefficients. These relations are the same as those we obtained in the previous section (\ref{relation1}-\ref{relation5}), where we imposed nonnegativity for the divergence of the entropy current.

\section*{Acknowledgements}

We would like to thank Synati Bhattacharya for valuble discutions. The work is supported in part by the Israeli Science Foundation center
of excellence.

\vskip 1cm

\appendix
\section{The entropy current divergence} \label{entropy div} 
Here we present the full computation of the entropy current divergence (\ref{divergence}):
\begin{align}\label{divergence2} 
\nabla_\mu J^\mu &= \frac{2\eta s}{p}\sigma_{\alpha \beta}\sigma^{\alpha \beta} +  (a_1 + \frac{s\zeta}{p})D^2 \text{ln} p + \frac{da_1}{dp} p(D\text{ln} p)^2 +  \frac{da_2}{dp} p a^\mu \nabla_\mu \text{ln} p + a_2 (\sigma^2-\omega^2 + R_{00}) \nonumber\\
&+\left(p\frac{dA_6}{dp} -p \frac{dA_3}{dp} +2B_4 -p\frac{dB_5}{dp} +B_6-\frac{de_2}{dp}\right)D\text{ln} p R_{00} \nonumber\\
&+\left(p \frac{dA_5}{dp} - \frac{p}{2} \frac{dA_3}{dp}-\frac{de_1}{dp} \right)RD\text{ln} p \nonumber\\
&+ \left( A_3 - p^{-2}\kappa_1\right) R^{\mu\nu}\sigma_{\mu\nu} -  \left(A_3+ p\frac{dA_3}{dp}- p\frac{dB_5}{dp}\right) R^{\mu\nu}a_\nu u_\mu \nonumber \\
& + \left(p\frac{dA_4}{dp} + 2B_2 - 2p\frac{dB_5}{dp} - \frac{2d_3}{p} -\frac{dd_4}{dp}\right) D^2\text{ln} pD\text{ln} p \nonumber\\
& + \left(-p\frac{dB_4}{dp}-2B_4 + p^2\frac{d^2B_5}{dp^2}+p\frac{dB_5}{dp}-p\frac{dB_6}{dp}+\frac{2d_5}{p}-\frac{dd_5}{dp} \right) \mathfrak{a}^2D\text{ln} p  \nonumber\\
&+\left( A_4-\frac{d_4}{p}\right)D^3\text{ln} p+ \left( -p\frac{dB_5}{dp}+B_6+\frac{2d_5}{p}\right) \mathfrak{a}^\beta \nabla_\beta(D\text{ln} p) \nonumber\\
&+\left( +2B_3-2B_4-p^2\frac{d^2B_5}{dp^2}-p\frac{dB_7}{dp}-\frac{2d_1}{p}+\frac{2d_5}{p} - p^{-2}\lambda_4 \right)\sigma^{\beta \mu}\nabla_\mu \text{ln} p \nabla_\beta \text{ln} p \nonumber \\
& + \left(A_5-\frac{e_1}{p} \right)DR+ \left(A_6-\frac{e_2}{p} \right)
 DR_{00}\nonumber\\ 
&+\left(p\frac{dB_1}{dp} - 2B_1 +2B_4-2p\frac{dB_5}{dp}+B_6+\frac{dd_2}{dp}-\frac{2d_2}{p}\right)\omega^2D \text{ln} p   \nonumber \\
&+ \left(p\frac{dB_3}{dp} - 2B_3 + 2B_4-2p\frac{dB_5}{dp}+B_6-B_7-\frac{dd_1}{dp}+\frac{2d_1}{p} - p^{-2}\lambda_0\right)\sigma^2D \text{ln} p  \nonumber\\
&+\left(-p^{-2}\lambda_1- 2B_3+\frac{2d_1}{p}\right)\sigma^{\nu \lambda} \sigma_{\nu \sigma}\sigma^\sigma_\lambda \nonumber \\
&+\left(-p^{-2}\kappa_2- 2B_3+\frac{2d_1}{p}\right)\sigma^{\nu \lambda} K_{\nu \lambda} \nonumber\\
&- \left(2B_3+4B_1+\frac{4d_2}{p} +p^{-2}\lambda_3-\frac{2d_1}{p}\right)\sigma^{\nu \lambda}\omega_\lambda^{~ \sigma}\omega_{\sigma \nu}   \nonumber \\
&-\left(p^2\frac{d^2B_5}{dp^2}+p\frac{dB_5}{dp} -p\frac{dB_2}{dp}+\frac{dd_3}{dp}\right)(D\text{ln} p)^3 \nonumber\\
& -\left(2B_3+p\frac{dB_5}{dp}+B_7-\frac{2d_1}{p}+p^{-2}\tilde{\tau}\right)\sigma^{\mu\nu}\nabla_\mu\nabla_\nu \text{ln} p - B_7 \mathfrak{a}_\nu \nabla_\mu \sigma^{\mu\nu}
\end{align}

\section{Curved background solution}\label{Curved background solution}
In this appendix, we explain briefly how to get the results of the stress tensor and entropy current for a curved hypersurface . 
The procedure is similar to that used in  \cite{Compere:2011dx,Compere:2012mt,Eling:2012ni} for a flat hypersurface. We solve the Einstein equations: $ \bf R_{AB}=0$, and boost the solution with the velocity vector $u^\mu$. Then, we promote the parameters of the solution to be depended on space-time coordinates of the cut off hyper surface located on $r=r_c$. Thus, we obtain the solution:  
\begin{align}
ds^2 = -(1+p^2(x^\lambda)(r-r_c)) u_\mu(x^\lambda) u_\nu(x^\lambda) dx^\mu dx^\nu - 2 p(x^\lambda) u_\mu(x^\lambda) dx^\mu dr + P_{\mu \nu}(x^\lambda) dx^\mu dx^\nu \ . \label{relativ1}
\end{align}
where we choose the gauge choice to be:
\begin{align}
g_{rr} = 0; \quad g_{r \mu} = -p u_\mu \ .
\end{align}
Since we promote the parameters to be dependent on the coordinates of space-time, the metric  does not satisfy the Einstein equations. We will
 solve the Einstein equations order by order in the derivative expansion. The Einstein equations at the $\text{n}^{\text{th}}$ order are:
\begin{align}
\bf \delta R^{(n)}_{AB} + \hat{R}^{(n)}_{AB} = 0  \ ,
\end{align}
where $ \bf \hat{R}^{(n)}_{AB} $ is the $\text{n}^\text{th}$ order Ricci tensor calculated from the $\text{(n-1)}^\text{th}$ order metric, and $\bf \delta R^{(n)}_{AB} $ is calculated from the correction of the metric to the $\text{n}^\text{th}$ order. The final result is
\begin{align}
\bf \delta R^{(n)}_{rr} &=  - \frac{1}{2} \partial^2_r (P^{\lambda \sigma} \delta g^{(n)}_{\lambda \sigma})\\
\bf \delta R^{(n)}_{r \mu} &= \frac{1}{4} p u_\mu \partial_r (P^{\lambda \sigma} \delta g^{(n)}_{\lambda \sigma}) + \frac{1}{2} p^{-1} \partial^2_r( u^\lambda \delta g^{(n)}_{\mu \lambda})\\
\bf \delta R^{(n)}_{\mu \nu} &= -\frac{1}{2} \left( u_\mu \partial_r (u^\lambda \delta g^{(n)}_{\nu \lambda}) + u_\nu \partial_r (u^\lambda \delta g^{(n)}_{\mu \lambda}))\right) - \frac{1}{2} \partial_r (\delta g^{(n)}_{\mu \nu}) - \frac{1}{2} p^{-2} \Phi  \partial^2_r (\delta g^{(n)}_{\mu \nu}) \nonumber \\ & - \frac{1}{2} u_\mu u_\nu \partial_r (u^\lambda u^\sigma \delta g^{(n)}_{\lambda \sigma}) + \frac{1}{4} \Phi u_\mu u_\nu \partial_r (P^{\lambda \sigma} \delta g^{(n)}_{\lambda \sigma}).
\end{align}
where $\Phi = 1+p^2(r-r_c)$.\\
This solution needs to be consistent with the following boundary conditions: (i) no singularity at $r=0$ and (ii) a given curved induced metric on $r=r_c$. The second implies that all the $n \geq 1$ order corrections vanish at $r=r_c$. Projecting into components normal and transverse to $u^\mu$ we find,
\begin{align}
P_\mu^\lambda P_\nu^\sigma \delta g^{(n)}_{\lambda \sigma} &= 2 p^2 \int^{r}_{r_c} \frac{1}{\Phi} dr' \int^{r'}_{r_c - \frac{1}{p^2}} P_\mu^\lambda P_\nu^\sigma \hat{\bf{R}}^{(n)} _{\lambda \sigma} dr'' \label{gensol1} \ , \\
u^\lambda P^\sigma_\mu \delta g^{(n)}_{\lambda \sigma} &= (1/2) (1-r/r_c) V^{(n)}_\mu(x) - 2 p \int^{r_c}_{r} dr' \int^{r_c}_{r'} dr'' P^\lambda_\mu   \hat{\bf{R}}^{(n)}_{r \lambda} \label{gensol2} \ , \\
u^\lambda u^\sigma \delta g^{(n)}_{\lambda \sigma} &= (1-r/r_c) A^{(n)}(x) + p \int^{r_c}_{r} dr' \int^{r_c}_{r'} dr'' \left(p P^{\lambda \sigma} \hat{ \bf{R}}^{(n)}_{\lambda \sigma} - p^{-1} \Phi \hat{\bf{R}}^{(n)}_{rr} - 2 \hat{\bf{R}}^{(n)}_{r \lambda} u^\lambda \right). \label{gensol3}
\end{align}
where $V^{(n)}_\mu$ ($V^{(n)}_\mu u^\mu = 0$) and $A^{(n)}$ are free, undetermined functions at this stage.
To determine these functions we calculate the Brown-York stress-energy tensor on the hyper surface located at $r = r_c$. We impose a frame in which the pressure does not receive any derivative corrections. This condition eliminates the derivative parts of the term proportional to $g_{\mu \nu}$ in 
the Brown-York stress-energy tensor and fixes $A^{(2)}$ to a non-zero value. Note, that the stress tensor in this frame has a term proportional to $u_\mu u_\nu$. Hence, at this point we see the traditional Landau frame $T_{\mu \nu} u^\mu = 0$ will be inconsistent. To fix $V^{(2)}_\mu$ we instead require $T^{\mu\nu}_{(n)}P^\lambda_\mu u_\nu = 0$.

With these values fixed, the stress-energy tensor can be put in a more conventional form:
\begin{align}
T_{\mu \nu} &=pP_{\mu \nu} -2 \sigma_{\mu\nu} - 2 p^{-1} u_\mu u_\nu (\sigma_{\alpha \beta}  \sigma^{\alpha \beta} + R) - 2p^{-1} \sigma_{\mu \rho} \sigma^\rho_\nu  - 4 p^{-1} \sigma^\rho_{(\mu}  \omega_{|\rho|\nu)}  - 4 p^{-1} \omega_{\mu \rho} \omega^\rho_{\hspace{3pt} \nu}
\nonumber \\
&  - 4 p^{-1} P^\alpha_\mu P^\beta_\nu \nabla_\alpha \nabla_\beta \text{ln} p  - 4 p^{-1} \sigma_{\mu \nu} D \text{ln} p  + 4 p^{-1} D^\perp_\mu \text{ln} p D^\perp_\nu \text{ln} p  - 2 p^{-1}P^\alpha_\mu P^\beta_\nu R_{\alpha\beta}  - 2 p^{-1} K_{\mu\nu}
\end{align}
Due to the $u_\mu u_\nu$ term, one can see that the energy density is no longer zero. It is corrected at the second order:
\begin{align}
\epsilon &= T_{\mu \nu} u^\mu u^\nu = -\frac{2}{p} \sigma_{\mu \nu}  \sigma^{\mu \nu} -\frac{2}{p}R \ .
\end{align}
The corrections due to the curved hypersurface metric alone are presented in (\ref{curvedmetricC}).

Since metric solution is no longer stationary, the event horizon is dynamical and its location varies in time and space. In order to find $r_h(x^\mu)$  we will need to solve the following equation in the derivative expansion
\begin{align}
g^{A B}  \partial_A (r-r_h(x^\mu)) \partial_B (r-r_h(x^\mu)) = 0 \ .
\end{align}
Using our previous results for the metric, it is straightforward to show at second order
\begin{align}
r_h &= r_c - \frac{1}{p^2} +  \frac{2}{p^3} u^\mu \nabla_\mu \text {ln} p - \frac{3}{2p^4} \sigma_{\alpha \beta} \sigma^{\alpha \beta} - \frac{1}{2p^4} \omega_{\alpha \beta} \omega^{\alpha \beta}  - \frac{8}{p^4} D \text {ln} p  D \text {ln} p  \nonumber \\
&+ \frac{1}{p^4} D^{\perp \mu} \text {ln} p  D^\perp_ \mu  \text {ln} p + \frac{4}{p^4} D (D \text {ln} p) - \frac{3}{2p^4} \left(R-R_{00}\right)\ .
\end{align}
Now, it is easy to construct  the normal to the event horizon by $\ell^A = g^{A B} \partial_B (r - r_h)$ and the metric determinant.
We find the entropy current
\begin{align}
J^\mu &= \frac{\ell^\mu}{4G} \left(1 - \frac{1}{p^2} \left(\sigma_{\alpha \beta} \sigma^{\alpha \beta}  - \frac{5}{2} \omega_{\alpha \beta}  \omega^{\alpha \beta} + 2 P^{\alpha \beta} \nabla_\alpha \nabla_\beta \text{ln} p  - 2 P^{\alpha \beta} \nabla_\alpha \text{ln} p \nabla_\beta \text{ln} p + R + 2R_{00} \right) \right) \ .
\end{align}

\end{document}